\documentclass[graybox]{svmult}

\usepackage{orcidlink}
\usepackage{rotating}
\usepackage{caption}
\usepackage{tikz}
\usetikzlibrary{positioning}
\usepackage{pst-all}
\usepackage{epsfig}
\usepackage{pdfpages}

\usepackage[dvips]{psfrag}

\usepackage{pgfplots}
\pgfplotsset{compat=newest}
\usetikzlibrary{intersections}
\usepackage{mathtools}

\usepackage{type1cm}        % activate if the above 3 fonts are
\usepackage{makeidx}         % allows index generation
\usepackage{graphicx}        % standard LaTeX graphics tool
\usepackage{multicol}        % used for the two-column index
\usepackage[bottom]{footmisc}% places footnotes at page bottom

\usepackage{newtxtext}       % 
\usepackage[varvw]{newtxmath}       % selects Times Roman as basic font

\newcommand{\be}{\begin{equation}}
  \newcommand{\ee}{\end{equation}}
\newcommand{\bes}{\begin{equation*}}
  \newcommand{\ees}{\end{equation*}}
\newcommand{\bea}{\begin{eqnarray}}
\newcommand{\eea}{\end{eqnarray}}
 \newcommand{\beas}{\begin{eqnarray*}}
\newcommand{\eeas}{\end{eqnarray*}}                 
\newcommand{\lb}{\label}
\newcommand{\bdm}{\begin{displaymath}}
\newcommand{\edm}{\end{displaymath}}

\makeindex             % used for the subject index
\begin{document}

\title*{From a quantum world to our classical Universe}

\author{Claus Kiefer\orcidlink{0000-0001-6163-9519}}

  \institute{Claus Kiefer \at University of Cologne, Faculty of
    Mathematics and Natural Sciences, Institute for Theoretical
    Physics, Cologne, Germany,
    \email{kiefer@thp.uni-koeln.de}} 

  \maketitle

\abstract {Modern cosmological theories invoke the idea that all
  structure in the Universe originates from quantum fluctuations. Understanding the
  quantum-to-classical transition for these fluctuations is of central
importance not only for the foundations of quantum theory, but also
for observational astronomy. In my contribution, I review the
essential features of this transition, emphasizing in particular the
role of Alexei Starobinsky.}

\section{Introduction}
\label{Intro}

Astronomical observations show the presence of structure in the
Universe. Visible and dark matter are grouped into galaxies and
clusters of galaxies. An impressive example is displayed in
Fig.~1. The photo was made in 12.5 hours of exposure with the James
Webb Space Telescope. It shows members of a distant galaxy cluster at
about 4.6 billion lightyears away. In addition, one observes luminous
arcs that are generated by the dark matter
dominated mass of this 
cluster via gravitational lensing and that originate from a much more distant
galaxy at about 9.5 billion lightyears. 

\begin{figure}[h]
  \begin{center}
    \includegraphics[width=0.8\textwidth]{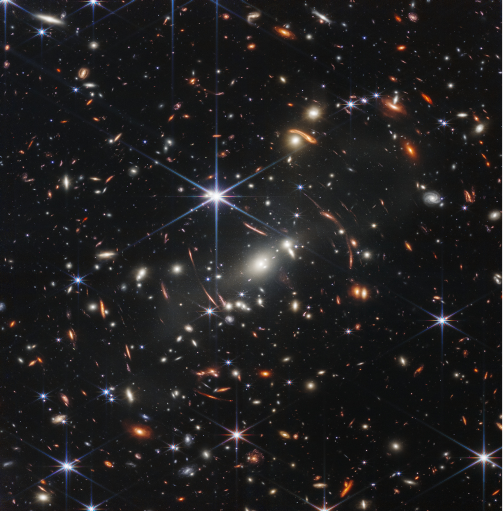}
  \caption[]{{\bf Webb's First Deep Field}.  This view of the early
    Universe was achieved with the NIRCam instrument of the James Webb
    Spacer Telescope in 2022. \\ Image Credit: NASA, ESA,
    CSA, STcI, NIRCam.}  
\end{center}
\end{figure}

An understanding of the origin of structure must be based on 
an understanding of gravity. In fact, much work was and is devoted to doing
(mostly numerical) calculations using general relativity or its
Newtonian limit in order to understand the evolution of structure from
initial conditons.

But what are these initial conditions? The perhaps most promising
candidate for a theory of the early Universe is {\em inflation}, the
idea that the Universe underwent a quasi-exponential expansion at a
very early time, but still much later than the Planck era. Alexei
Starobinsky, in the memory of whom this article is written, was one of the
pioneers of this idea.

A main feature of inflation is that structure has emerged from initial
{\em quantum} fluctuations of the spacetime metric and a
hypothetical inflaton field. But this raises the fundamental question
of how the transition from quantum to classical behaviour happened and
which physical, potentially observational, consequences this transition ensued. It
was this question that initiated my collaboration with Alexei and our colleague
David Polarski. The perhaps most exciting aspect in our work was (and
still is) the insight that an understanding of the foundations of
quantum theory and in particular the formation of entanglement between
quantum states is
needed to understand the origin of galaxies and galaxy clusters.  

In this article, I will focus on our work on the decoherence of the
primordial fluctuations. I have enjoyed working with Alexei also on
other topics, such as Higgs inflation and the cosmological constant,
which I will leave out here. 

My article is organized as follows. In Section~2, I will focus on the
quantum issues in the inflationary scenario and state the problem of
the quantum-to-classical transition. This problem is dealt with (and
solved, in my opinion) by invoking the process of decoherence, which
is familiar from ordinary quantum theory, where it was experimentally
confirmed. Section~3 gives a brief introduction to decoherence with
emphasis on its relevance for quantum cosmology. Section~4
then explains the decoherence of the primordial fluctuations and its
consequences for, in particular, their entropy. Finally, Section~5
focuses on the relevance of these investigations 
for the arrow of time in our Universe.

\section{The inflationary universe}

Alexei considered an exponential expansion of the early Universe
already in \cite{AS79}. He was mainly motivated by symmetry arguments and
explored the idea that the 
``universe was in a maximum symmetrical quantum state before the
beginning of the classical Friedman expansion''. This brief statement
involves the ideas that the Universe early on (before the
radiation-dominated phase) was in a quantum state
and that this state had maximum symmetry. Since he assumed this
maximum symmetry state to correspond to a spacetime solution different
from Minkowski space, he is led to considering a quantum state in de~Sitter space.

The only information from such a hypothetical early state, so the
argument, can come from gravitons (corresponding to gravitational
radiation in the semiclassical limit). Detection of such primordial
gravitons could even provide information about a fundamental physical
theory describing the symmetric state. Alexei concluded his paper with
the words:

\begin{quotation}
  It is remarkable that observation of the spectrum [of gravitons]
  would not only allow us to determine the initial state of the
  universe, but also to measure experimentally the $s$
  constant\footnote{In \cite{AS79}, this constant is defined as $s=H_0l_{\rm
      P}$, where $H_0$ is the Hubble parameter of inflation, and
    $l_{\rm P}$ is the Planck length.} which
  is fundamental for a unified theory of interactions.
\end{quotation}

So far, primordial gravitons have not been detected, but their
observation at least by indirect means belongs to the central goals of
work on inflation. The primordial gravitons, together with the
primordial density fluctuations, also play a central role in our
research on decoherence reviewed below.

Ideas about an initial state of maximum symmetry are, in fact, old and
can be traced back to the pre-Socratics. Anaximander, who lived in the
sixth century BC in Miletus, postulated the {\em apeiron} as the
source of all things.\footnote{Cf. the remarks in \cite{CZ91}, in
  particular footnote~1.} He also
used symmetry arguments to conclude that the Earth must be round and freely
floaring in empty space.\footnote{Cf. the excellent discussion in
  \cite{Popper98}.} Anaxagoras, another Greek philosopher, who lived in
the fifth century BC 
partly in Athens, entertained the idea of a homogeneous initial state
out of which all constituents of the later cosmos would emerge. 

To implement his ideas of maximum symmetry, Alexei considered a model
of exponential expansion that is based on
one-loop quantum-gravitational corrections to the Einstein field
equations \cite{AS80}. This model is now
called {\em Starobinsky inflation} and constitutes one of the best-fit
models to observations of the \textsc{Planck} satellite and other
missions. A selection by observation is of utmost importance, since
here exist many models of inflation, even if one restricts 
attention to the observationally preferred slow-roll single field
models \cite{encyclopaedia}.

A major issue in inflation is the calculation of the power spectrum
for scalar and tensor fluctuations in de~Sitter space; see, for
example, \cite{Baumann12} for an extensive review and
references. Besides his already mentioned work on the tensor
(graviton) fluctuations, Alexei also presented an early calculation of
the adiabatic amplitude of adiabatic scalar fluctuations generated
during inflation \cite{AS82}.

The standard picture for the evolution of the fluctuations is shown in
Fig.~2 and explained in the caption. The essential point is that a
causal evolution is possible -- a fluctuation with a certain
wavelength $\lambda$ starts from within the Hubble scale during
inflation, leaves the Hubble scale during this phase and later
re-enters the Hubble scale during the radiation (or matter-) dominated
phase. The power spectra after re-entry then serve as initial
conditions for the formation of structure as seen, for example,
in Fig.~1. 

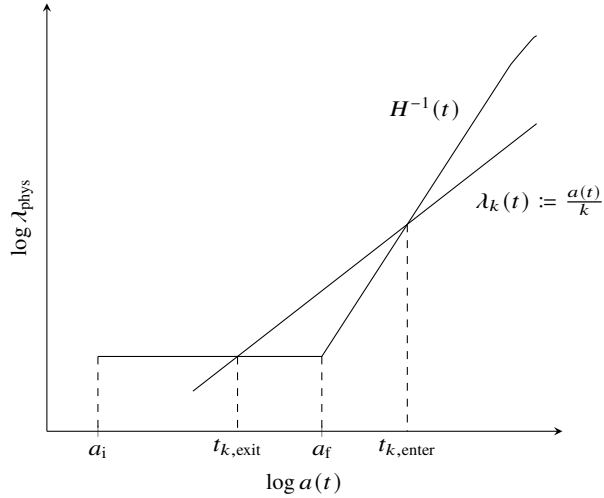
\begin{figure}[h]
\begin{center}
  \begin{tikzpicture}

\begin{loglogaxis}[
    axis x line = bottom,
    axis y line = left,
    xlabel = $\log a(t)$,
    ylabel = $\log \lambda_\text{phys}$,
    xmin = 1e-57,
    xmax = 1e3,
    ymin = 1e-61,
    ymax = 1e3,
    xtick = {8.756510762696519e-52, 9.999999999999999e-26},
    xticklabels={$a_\text{i}$, $a_\text{f}$},
    ytick = \empty,
    clip = false,
    ]
\addplot [no markers,
    name path = h]
    table
        {scaleDat.txt}
    node[above left,
        pos = .2]
        {$H^{-1}(t)$};
\addplot [no markers,
    name path = lambda]
    table {
        1e-0 1e-15
        1e-40 1e-55
        }
    node[below right,
        pos = .2]
        {$\lambda_k(t) \coloneqq \frac{a(t)}{k}$};

% https://tex.stackexchange.com/questions/91310/tikz-difference-between-node-and-coordinate
\coordinate (ai) at (8.756510762696519e-52, 1.54603595933673e-50); 
\coordinate (af) at (9.999999999999999e-26, 1.54603595933673e-50); 

% https://tex.stackexchange.com/questions/449138/how-to-use-intersections-in-pgfplots-as-x-ticks
% https://tex.stackexchange.com/questions/312597/pgfplots-macro-to-draw-lines-from-point-to-y-axis

\draw [dashed,
    name intersections={of=h and lambda,name=its}
    ]
        (ai) -- (ai |- {rel axis cs:0,0})
        (af) -- (af |- {rel axis cs:0,0})
        (its-1) -- (its-1 |-{rel axis cs:0,0})
            node[below] {$t_{k, \text{enter}}$}
        (its-2) -- (its-2 |-{rel axis cs:0,0})
             node[below] {$t_{k, \text{exit}}$};

\end{loglogaxis}

\end{tikzpicture}
\caption[]{Time development of a physical scale
  $\lambda_k(t):= a(t)/k$, where $a(t)$ is the scale factor of a
 Friedmann universe and $k$ is the dimensionless wave number, and the
  Hubble scale $H^{-1}(t)$. During an 
  inflationary phase, $H^{-1}(t)$ remains approximately constant.
  After the end of inflation
  ($a_{\rm f}$), the Hubble scale $H^{-1}(t)$ increases faster than any
  physical scale.
  Therefore the scale described by $\lambda_k$, which has left the
  Hubble scale at time $t_{k,{\rm exit}}$, enters the Hubble scale
  again at $t_{k,{\rm enter}}$ 
  in the radiation- (or matter-) dominated phase. This figure was
  produced by using real data from the \textsc{Planck} satellite and using the
  astropy package in python, see www.astropy.org/acknowledging.\\ I am
  grateful to Yi-Fan Wang for producing this figure. Figure reprinted
from \cite{OUP4}. \copyright~C. Kiefer}
\end{center}
\end{figure}

How do we describe the emergence and evolution of quantum fluctuations
during an inflationary phase? From a fundamental point of view, we
should start with a quantum theory of gravity \cite{OUP4}. Such a
theory is not yet available in complete form, but there exist various
approaches on which a discussion can be based. One of the most
suitable approach for our purpose is quantum geometrodynamics, with the
Wheeler--DeWitt equation playing the central role \cite{OUP4,CK09}. In
quantum geometrodynamics, it is possible to perform a consistent
semiclassical approximation scheme, which is sufficient for the
treatment of the inflationary fluctuations. Symbolically, one can
write the total quantum state of gravity $\Psi$, which is a solution
of the Wheeler--DeWitt equation, in the form
\be
\label{semiclassicalstate}
\Psi\propto \exp({\mathrm i}
  m_{\mathrm{P}}^2S_0[\mathrm{gravity}]) 
  \psi[\mathrm{gravity}, \mathrm{matter}],
  \ee
where $m_{\mathrm{P}}$ denotes the Planck mass. Performing an
expansion with respect to the Planck mass, one finds that $S_0$ obeys
the Hamilton--Jacobi equation of gravity (providing a semiclassical
background allowing, in particular, the introduction of a time parameter) and that
$\psi$ satisfies a (functional) Schr\"odinger equation for 
quantum fields on that background \cite{OUP4,Chataignier22}. Higher orders in the
Planck-mass expansion yield quantum-gravitational correction terms to
the Schr\"odinger equation \cite{KS91}. For the discussion below, we
will not need these correction terms; they are important for searches
of genuine quantum-gravitational effects in cosmology such as the
modification of the CMB anisotropy spectrum at large scales
\cite{Chataignier22,LKM23}.

The quantum fluctuations in inflation can be described by the
``Mukhanov--Sasaki variable'', which is a gauge-invariant combination
of the scale factor $a$, the homogeneous part of the inflaton field, and their 
perturbations \cite{Baumann12}. For our discussion of decoherence, it
is not necessary to employ the full gauge-invariant formalism; it is
sufficient to describe them by a massless scalar field $\phi$ in a
Friedmann--Lema\^{\i}tre universe (here chosen to be spatially
flat). This formalism can encompass both scalar and tensor
modes. Since the fluctuations are small, there mutual interaction is
negligible, so they can be treated as being independent of each other and
can be characterized by their (dimensionless) wave number $k$ (absolute value of
the vector $\mathbf k$) or their wavelength $\lambda_k=a/k$. Because of this, it is
appropriate to go to Fourier variables, 
\be
\lb{yk}
y_k:= a\phi_k=y_{-k}^*.
\ee
The Fourier modes are complex, but for our purpose it is sufficient to
focus attention on one real mode $y_k$ (in the following skipping the
index $k$). 

In the semiclassical approximation to quantum gravity, each mode can
be characterized by a wave function $\psi$ obeying a
quantum-mechanical Schr\"odinger equation separately for each mode,
\be
\lb{Schroedinger}
{\rm i}\hbar\frac{\partial\psi(y,t)}{\partial t}=\hat{H}\psi(y,t),
\ee
where the Hamiltonian $\hat{H}$ is given by
\be
\lb{Hamiltonian}
\hat{H}=\frac12\left(p^2+k^2y^2+\frac{2\dot{a}}{a}yp\right);
\ee
$p$ denotes the momentum conjugate to $y$.
The first two terms are standard harmonic oscillator terms, while the
last term describes the coupling of the mode to the evolving universe
described by $a(t)$. It is well known from quantum optics that this
last term leads to a squeezing of the quantum state (characterized by increasing
variance of either $y$ or $p$ and decreasing variance of the
respective other variable). The squeezing formalism for the cosmic
perturbations was first presented in \cite{GS90}. Taking into account
both the real and the imaginary part of $y_k$ in \eqref{yk} leads to a
two-mode squeezed state.

In order to solve the Schr\"odinger equation \eqref{Schroedinger} for
the modes, we need to impose an initial condition. This may be
straightforward to do in quantum optics, but is far from trivial in
cosmology. The simplest, and most commonly used, initial condition is
the adiabatic vacuum state already used by Alexei in \cite{AS79}. In
the Schr\"odinger picture, it is given by the mode wave function
\be
\lb{initial}
\psi_0(y)=\left(\frac{2k}{\pi}\right)^{1/4}
\exp\left(-ky^2\right).
\ee
This corresponds to the usual harmonic oscillator ground state.
With this initial state and the Hamiltonian
\eqref{Hamiltonian}, the solution of \eqref{Schroedinger} remains of
Gaussian form and can be written as
\be
\lb{solution}
\psi(y,t)=\left(\frac{2\Omega_{\rm R}(t)}{\pi}\right)^{1/4}
\exp\left(-\Omega(t)y^2\right)\ , \quad
\Omega:=\Omega_{\rm R}+\I\Omega_{\rm I}.
\ee
The dynamics leads to a squeezing of the vacuum state and can be
parametrized as
\be
\lb{parameters}
\Omega_{\rm R}=\frac{k}{\cosh
2r+\cos2\varphi\sinh 2r}\ , \quad \Omega_{\rm I}=-\Omega_{\rm
R}\sin2\varphi\sinh2r ,
\ee
with $\Omega_{\rm R}=k$ and $\Omega_{\rm I}=0$ for the initial
state. Here, $r$ is called squeezing parameter, and $\varphi$ is
called squeezing angle. Figure~3 shows the variances (resp. contours
of the Wigner function) for the initial vacuum state (blue circle) and
the resulting squeezed vacuum state.

For eternal inflation, one finds asymptotically $r\to\infty$ and
$\varphi\to 0$, that is, the major axis with size ${\rm
  e}^r\to\infty$ of the ellipse becomes aligned 
to the $y$-axis, and the minor axis becomes arbitrarily small. Since,
however, inflation is not eternal, this limit is not obtained, and we
are left with a finite, but large value of the squeezing parameter
$r$.

\begin{figure}[h]
  \begin{center}
    \includegraphics[width=0.6\textwidth]{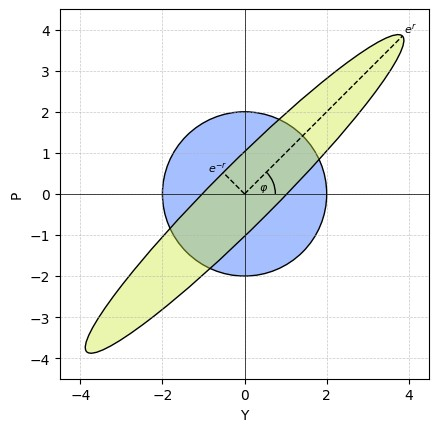}
  \caption[]{Squeezed vacuum state (ellipse in green) compared with the vacuum
  state (circle in blue).}  
\end{center}
\end{figure}

For concrete cases of evolution laws $a(t)$, the parameters can be
given explicitly (at least numerically). In the simplest case of
de~Sitter exponential evolution,
\bdm
a(t)=a_0\exp(H_{\rm I}t),
\edm
one finds \cite{PS96}
\bdm
\sinh r=\frac{aH_{\rm I}}{2k}\ , \quad\cos2\varphi=\tanh r .
\edm
For $r\gg 1$, this gives ${\rm e}^r\approx aH_{\rm I}/k$, so the
squeezing parameter scales approximately with the logarithm of the
scsale factor. For the largest cosmological scales, one has $r\approx
120$ and $\varphi\approx 0$. In this limit, the major axis of the ellipse seen in
Fig.~3 will then be aligned along the $y$-axis. 

So far, we have seen that the primordial fluctuations are in a (two-mode)
squeezed quantum state. But how, then, can one understand the
transition to a state describing classical stochastic fluctuations as
they are used when analysing the CMB anisotropy spectrum and needed to
understand the structure shown in Fig.~1? 

A first step in this direction was achieved in the seminal paper 
\cite{PS96}. There it was shown that quantum expectation values
calculated from \eqref{solution} cannot be distinguished from
classical stochastic expectation values in the limit $r\to\infty$
(corresponding to neglecting what is called, in the Heisenberg
picture, the ``decaying mode'').
For illustration, the analogy of a simple quantum-mechanical
example is presented in \cite{KP98}.

But this is not yet enough. As is well known, squeezed states are very
sensitive to interactions with other degrees of freedom (their
`natural environment') \cite{deco}. This feature is especially
pronounced for states with a broad variance in position (this role is here played
by the variable $y$), because most realistic interactions are in
position, not in momentum. This leads to the process of decoherence,
which decides which variables do assume classical properties and which
do not. Decoherence as applied to the squeezed quantum state
\eqref{solution} was first discussed in \cite{KPS98}, but before we
turn to this, we present a brief history of decoherence with an eye
towards applications in quantum cosmology.

\section{A brief history of decoherence}
\label{decoherence}

In 1927, the formalism of quantum theory as it is used today was more
or less complete. A central role therein is played by states
(wave functions) obeying the superposition principle. The same
year 1927 also witnessed the beginning of the still ongoing debate about the
correct interpretation of the 
theory and, in particular, about the meaning of the wave functions. In 1929,
Nevill Mott addressed the important question why a particle described
by a spherically-symmetric wave function does not lead to a spherical
structure in a cloud 
chamber, but instead to a trajectory \cite{Mott29}. To cite from that paper:

\begin{quotation}
In the theory of radioactive disintegration, as presented by Gamow,
the $\alpha$-particle is represented by a spherical wave which slowly
leaks out of the nucleus. On the other hand, the $\alpha$-particle,
once emerged, has particle-like properties, the most striking being
the ray tracks that it forms in a Wilson cloud chamber. It is a little
difficult to picture how it is that an outgoing spherical wave can
produce a straight line. 
\end{quotation}

Mott's important insight was the observation that the quantum nature
of the many atoms in the cloud chamber must be taken adequately into account when
dealing with their interaction with the $\alpha$-particle. In this, it
is of the utmost importance to note that the total wave function is
not a wave function in ordinary three-dimensional space (as
e.g. classical fields are), but in the high-dimensional configuration
space of $\alpha$-particle {\em plus} cloud chamber atoms.
In Mott's words:

\begin{quotation}
The difficulty that we have in picturing how it is that a spherical
wave can produce a straight track arises from our tendency to picture
the wave as existing in ordinary three dimensional space, whereas we
are really dealing with wave functions in the multispace formed by the
co-ordinates both of the particle and of every atom in the Wilson chamber.
  \end{quotation}

  Mott then went on to show that the full quantum state can be written
  as a superposition of track-like states for the $\alpha$-particle,
  with each of these states correlated with a corresponding state of
  the chamber atoms. In todays's language (which goes back to
  Schr\"odinger's work in the 1930s) this total state is called an
  {\em entangled state}. Mott then argued (as was common in those days
  and happens sometimes even today) that this superposition can be
  replaced by a statistical ensemble with Born probabilities for the
  various components. What is observed in the cloud chamber is then one
  component of this total state with a probability given by the Born rule.

  From a heuristic and pragmatic point of view, such an attitude may be
  understandable, but it is in conflict with the formalism of
  quantum theory. A superposition of states is fundamentally and
  observationally different
  from a statistical {\em ensemble} as myriads of interference experiments
  have demonstrated and continue to demonstrate. An electronic
  spin-state pointing in $x$-direction, for example, can be written as
  a superposition of spin states in $+z$- and $-z$-direction, but it
  is mathematically and experimentally different form an ensemble of
  spin states pointing into the $+z$- and $-z$-directions. Other
  examples can be found in particle physics where e.g. the superposition of
  a K-meson and its antiparticle results in a {\em new} particle
  (K-long or K-short) and not in a mixture of the original
  particles. The same holds for neutrino oscillations.

  So something important must be missing in Mott's analysis. The
  missing important ingredient is what is today called the process of {\em
    decoherence}. It was first discussed in a seminal paper by
  H.-Dieter Zeh in 1970 \cite{Zeh70}, not yet using the word
  decoherence. Zeh showed that it is
  inconsistent to treat macroscopic quantum systems as isolated systems
  because they become strongly entangled with their
  environment.\footnote{Here and below `environment' means degrees of
    freedom whose coupling to the system cannot be neglected, but
    which are not ``under control''. A
    standard example is a dust particle in interaction with air
    molecules or photons \cite{JZ85}.} These systems thus do {\em not}
  obey a Schr\"odinger equation, and their time evolution is {\em not} unitary.
  
  Why has it taken so long from
  Mott's to Zeh's paper? 
  The main reason is perhaps the prevalence of the Copenhagen
  interpretation of quantum theory after 1930 where classical concepts
  are postulated from the outset and where thus such questions
  were ignored. They were also considered irrelevant because
  quantum theory worked successfully at a pragmatic level; experiments
  able to test foundational issues were only possible after the advent
  of the laser and other tools. 

  The origin of Zeh's work lies in certain puzzling features of
  nuclear physics described by him, for example, in his essay
  \cite{Zeh06}. Since his line of reasoning reflects certains ideas
  that are relevant for quantum cosmology, I shall spend some
  words on it.

In nuclear physics, there are many examples where a many-nuclei system
is described by an entangled wave function $\Psi$ obeying the
stationary Schr\"odinger equation
  \be
  \lb{HPsiEPsi}
  H\Psi=E\Psi.
  \ee
Nevertheless, for a calculation of the spectra there exist successful
methods that make use of time-{\em dependent} single-nucleon wave
functions (e.g. using the `Hartree--Fock approximation'). These spectra also
show features of deformed nuclei (e.g. Coriolis-force type of
effects), in spite of the total state $\Psi$ 
in \eqref{HPsiEPsi} being in an angular momentum eigenstate. In
fact, the state $\Psi$ can be written as a superposition of states
with different orientations in the form
  \be
  \lb{nuclei}
\Psi=\int{\mathrm D}\Omega\, f(\phi,\theta,\chi)U(\phi,\theta,\chi)\Phi({\mathbf
  x}_1, \ldots, {\mathbf x}_n),
\ee
where $U(\phi,\theta,\chi)$ denotes the unitary transformation
for a rotation with angles $\phi,\theta,\chi$ (e.g. the Euler angles),
${\mathrm D}\Omega$ is the volume element in this rotation space, and
$\Phi$ is, for example, the product of many asymmetric single-particle
wave functions $\phi_i({\mathbf x}_i)$ (or generally a determinant
wave function). The wave function $\Phi$ is then a `deformed' state,
which cannot be an eigenstate of \eqref{HPsiEPsi}, because such eigenstates
are rotation-invariant angular-momentum eigenstates, exhibiting
rotational symmetry instead of definite orientation. Only the
superposition over all deformed nuclei in the form of \eqref{nuclei} is
then an (approximate) solution to the Schr\"odinger equation
\eqref{HPsiEPsi}. This total state $\Psi$ is strongly {\em
  entangled} with respect to the many single-nuclei states.

This counterintuitive situation (full symmetric state from a superposition of
many states exhibiting a different direction) led Zeh in the 1960s to
a weird speculation. In his own words (\cite{Zeh06}, p.~4):

\begin{quotation}
So one may say that the individual nucleons ``observe'' an apparent
asymmetry in spite of the symmetric global superposition of all
orientations. \ldots This analogy led me to the weird speculation
about a nucleus that is big enough to contain a complex subsystem
which may resemble a registration device or even a conscious
observer. \ldots If the nucleons in the deformed nucleus dynamically
feel a definite orientation in spite of the global superposition,
would an internal observer then not similarly have to become ``aware
of'' a {\em certain} measurement result?
  \end{quotation}

  We know at least since the famous work of Einstein, Podolsky, and Rosen (EPR) 
  that entanglement is a central feature of quantum theory
  \cite{CK22}. It is of great importance to see that entanglement is
  not only responsible for genuine quantum aspects (as in the EPR
  gedanken experiment), but also for the classical {\em appearance} of
  our world. The key to understanding this is the above
  mentioned process of decoherence. Let me give a brief explanation
  (see \cite{deco} and \cite{memorial} for details).

  Following original ideas by John von Neumann, we consider an
  interaction between a system (S) and an apparatus (A), see
  Fig.~4. Both are described by quantum theory.
 
  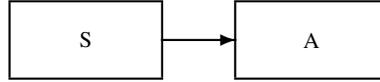
\begin{figure}[htb]
\begin{center}
\setlength{\unitlength}{1cm}
\begin{picture}(5,1) \thicklines %\linethickness{1mm}
\put(0,0){\framebox(2,1){S}}
\put(3,0){\framebox(2,1){A}}
\put(2,0.5){\vector(1,0){1}}
\end{picture}
\end{center}
\caption{Original form of the von Neumann measurement model.}
\end{figure}
  
  If the states of the measured system that are
to be distinguished by the apparatus are denoted by $|n\rangle$
(e.g. spin up and spin down), 
a simple interaction Hamiltonian can be formulated as follows:
\be H_{\rm int} =\sum_n|n\rangle\langle n| \otimes\hat{A}_n.
\label{Hamil} \ee
The operators $\hat{A}_n$ act on the states of the apparatus and 
must depend on the `quantum number' $n$ in order to describe a measurement.
Equation (\ref{Hamil}) describes an idealized situation in which
the apparatus becomes correlated with the system state without changing
the latter. There is thus no disturbance of the system by the
apparatus---on the contrary, the apparatus is disturbed by the system
(in order to yield a measurement result).

If the
measured system is initially in the state $|n\rangle$ and the apparatus in
some initial state $|\Phi_0\rangle$,
the evolution according to the Schr\"odinger equation
with the Hamiltonian (\ref{Hamil}) reads
\bea |n\rangle|\Phi_0\rangle \stackrel{t}{\longrightarrow}
     \exp\left(-{\rm i} H_{\rm int}t\right)|n\rangle|\Phi_0\rangle
     &=& |n\rangle\exp\left(-{\rm i}\hat{A}_nt\right)|\Phi_0\rangle\nonumber
\\
     &=: & |n\rangle|\Phi_n(t)\rangle.  \label{ideal} \eea
The resulting apparatus states $|\Phi_n(t)\rangle$ are called
{\em pointer states}.
A process analogous to (\ref{ideal}) can also be
formulated in classical physics. The essential quantum features
enter when one considers as initial state a {\em superposition} of different
eigenstates of the measured `observable' (such as a superposition of
spin up and spin down). The
linearity of time evolution then leads immediately to
\be \left(\sum_n c_n|n\rangle\right)|\Phi_0\rangle
    \stackrel{t}\longrightarrow\sum_n c_n|n\rangle
    |\Phi_n(t)\rangle. \label{measurement}
    \ee
 This state describes a superposition of macroscopic measurement results
(as in the gedanken experiment with Schr\"o\-ding\-er's cat). 
To avoid such a bizarre non-classical state, and to avoid the apparent conflict with
experience, von Neumann introduced a dynamical collapse of the
wave function as a new law that violates the universality of the
Schr\"odinger equation.
This alleged collapse should then select one component
with probability $\vert c_n\vert^2$ (according to the Born rule). 

The crucial idea to avoid such Schr\"odinger-cat type of states and to
arrive at a definite measurement result (more precisely, the
{\em appearance} of a definite result),
the unavoidable interaction with the environment must be taken into
account \cite{Zeh70}. Environmental degrees of freedom usually occur
in big numbers (e.g. in the case of scattering with photons or air
molecules \cite{JZ85}) and are not under control (Fig.~5). 

\begin{figure}[ht]
\begin{center}
\setlength{\unitlength}{1cm}
\begin{picture}(10,2)(-0.5,-0.5) \thicklines %\linethickness{1mm}
\put(0,0){\framebox(2,1){S}}
\put(3,0){\framebox(2,1){A}}
\put(2,0.5){\vector(1,0){1}}
\put(6,-0.5){\framebox(3,2){E}}
\put(5,0.3){\vector(1,0){1}}
\put(5,0.5){\vector(1,0){1}}
\put(5,0.7){\vector(1,0){1}}
\put(-0.5,-0.5){\dashbox{0.2}(6,2){}}
\end{picture}
\end{center}
\caption
{Realistic extension of the von Neumann
 measurement model including the environment. Classical properties 
emerge through the unavoidable, irreversible interaction
 of the apparatus with the environment.}
\end{figure}
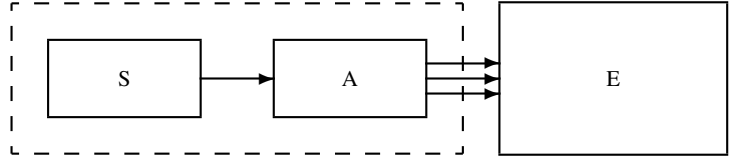

Assuming that the interaction of the apparatus with the environment
can again be described by a Hamiltonian of the form
(\ref{Hamil}), one arrives at a gigantic superposition
now {\em including} the environment, according to
\be \left( \sum_n c_n|n\rangle|\Phi_n\rangle\right)|E_0 \rangle
  \quad  \stackrel{t}\longrightarrow \quad
   \sum_n c_n|n\rangle |\Phi_n\rangle |E_n\rangle .\label{ideal2}
   \ee
 At first sight it seems as if the situation with macroscopic
 superpositions did even get worse. The important observation is,
 however, that most of the many
environmental degrees of freedom are inaccessible.
They thus have to be integrated out from the full state
(\ref{ideal2}) in order to describe what can be seen at the
apparatus. The result is a reduced density matrix 
for system plus apparatus, which contains all the information
that can be retrieved from the two,
\be
\rho_{\rm SA} \approx \sum_n |c_n|^2 |n\rangle\langle n|
   \otimes |\Phi_n \rangle \langle \Phi_n |
   \qquad\mbox{if}\qquad
   \langle E_n|E_m \rangle \approx \delta_{nm}.
 \label{deco}
\ee
In the last step we have used that, under realistic conditions,
different environmental states 
are orthogonal to each other. 
Equation~(\ref{deco}) is approximately identical to the
density matrix of an ensemble of measurement results
$|n\rangle |\Phi_n\rangle$. The system and apparatus thus seem to
be in one of the states $|n\rangle$ and $|\Phi_n\rangle$, given by
the probability $\vert c_n\vert^2$. This is also called an {\em
  apparent collapse}. In a universally valid quantum theory, the
interferences described by \eqref{deco} still exist -- they are
present in the entanglement between S and A with the environmental
degrees of freedom, but are no longer recognizable in S and A
alone. Or, in the words of \cite{JZ85}:
``The interferences exist, but they are not {\em there}.'' 

By now, the process of decoherence has been successfully tested in a
variety of experiments; see, for example, the corresponding
contributions to \cite{memorial}. Further developments in the general
formalism have led, for example, to the concept of 
{\em Quantum Darwinism} \cite{WZ22}.  This refers to the fact that
an abundance of many copies of the pointer
states (the classical states distinguished by the interaction with the
environment) are imprinted in the environment, so information about those
pointer states can be obtained from observing the environment. 

Decoherence originates from the entanglement between a system with its
environment. If gravitational degrees of freedom are also described in
quantum terms (as we assume here), they also take part in
decoherence. In an early simple model, it was shown how a quantum metric in
Newtonian approximation becomes decohered by the interaction with
gas molecules \cite{Joos86}. How decoherence can in principle be applied to a full
quantum theory of gravity and, in particular, to quantum cosmology,
was first suggested in \cite{Zeh86}. The first concrete calculations
in the context of the Wheeler--DeWitt equation of canonical quantum
gravity were presented in \cite{CK87}, see also  \cite{CK92}, where 
anlogous calculations are given for quantum electrodynamics. In these
papers, it was shown how the scale factor of an approximately
homogeneous and isotropy universe as well as the homogeneous part of
an inflaton are decohered by quantum fluctuations of gravity and matter.
To give one concrete example of a later calculation
\cite{BKKM99}, the reduced density matrix for the scale factor $a$
after integrating out the many degrees of freedom of primordial
gravitational waves was found to read
\be
\rho_0(a,a')\stackrel{t}{\to} \rho_0(a,a')\exp\left(-CH_{\rm I}^3a(a-a')^2\right)\ ,
\; C>0,
\ee
where $H_{\rm I}$ is the Hubble scale of inflation. The interpretation
of this result is that the homogeneous background, even if quantized,
assumes classical properties quickly during inflation, since
non-diagonal terms in the density matrix become tiny for increasing
$a$; see also the illuminating discussion in
\cite{BK22}.

While these quantum gravity considerations may seem mainly of conceptual
interest, the quantum-to-classical transition for the primordial
fluctuations in an inflationary universe is of more direct relevance
because it leads to observational consequences. It is here where our
collaboration starts \cite{KPS98}.

\section{Decoherence of primordial fluctuations}
\label{primordial}

As we have reviewed above, an initial adiabatic ground state for
metric and inflaton perturbations becomes highly squeezed during
inflation, see \eqref{solution} and \eqref{parameters}. The squeezing
appears in the momentum $p$ canonically conjugate to the field
amplitude $y$, that is, the resulting quantum state is very broad in
$y$. As is well known from quantum optics, broad states in
position are highly sensitive to an influence by environmental degrees
of freedom \cite{deco}. The reason is that usual interaction terms in
the Hamiltonian invoke a coupling in position, not in momentum. This
sensitivity motivated us in \cite{KPS98} to 
study possible decoherence effects arising from an interaction with
environmental degrees of freedom.

But where can such degrees of freedom come from? There are many
possibilities. The perhaps most natural one is to take into account
the interaction between modes of different wavelength, which arises
from the non-linear interaction in the full theory and which is
neglected in the approximation used in Sec.~2. Other options are
interactions with degrees of freedom from fundamental (as yet unknown)
theories that lead, for example, to string-effective actions containing
the coupling of many different fields. Fortunately, the details
of such interactions are not needed for understanding the major
features of 
decoherence. The only general input is the fact that interactions in
quantum field theory are ususally local interactions coupling field
amplitudes (and not their canonical momenta). With this, a quite
general formalism for the decoherence of primordial fluctuations can
be developed \cite{KPS00,KLPS07}.  

The simplest case is an ideal interaction in the spirit already discussed by von
Neumann. In this, there is not direct perturbation of the system, but
only the formation of entanglement between the system and the
interacting degrees of freedom. It can be described by the following
modification of the reduced density matrix:
\be
\lb{dm-decohered}
\rho_0(y,y')\longrightarrow
\rho_{\xi}(y,y')=\rho_0(y,y')
\exp\left(-\frac{\xi}{2}(y-y')^2\right),
\ee
where $\xi$ denotes a phenomenological parameter that contains the
details of the interaction and that can be calculated from the
parameters and the form of a given interaction Hamiltonian. 

In order to describe a realistic decoherence condition for the
primordial fluctuations, one has to assume that $\xi$ is much bigger
than the parameter $\Omega_{\rm R}$ appearing in \eqref{solution} and
given in \eqref{parameters},
\bdm
\xi\gg \Omega_{\rm R}\approx k\E^{-2r}.
\edm
Otherwise, \eqref{dm-decohered} does not describe a decrease in
off-diagonal density-matrix terms, as is required for decoherence.
It turns out that the
decoherence time scale $t_{\rm d}$ for the modes during inflation is (apart from
prefactors) given by the Hubble scale $H_{\rm I}^{-1}$ of inflation,
$t_{\rm d}\sim H_{\rm I}^{-1}$ \cite{KPS00}. Although the system describing the
modes is not a chaotic system (but only classically unstable), there
is nevertheless a close analogy to a chaotic system: the Hubble
parameter $H_{\rm I}$ is analogous to the Lyapunov parameter. The
decoherence time can also be calculated from the rate of de-separation,
that is, the rate at which originally unentangled states become
entangled \cite{KP98}.

Correlations between $y$ and $p$ can easily been recognized from
the contour of the Wigner function (the Wigner ellipse). In the case
of Gaussian systems, as is the case here, the
Wigner function remains positive, in spite of the presence of quantum
features. If the modes are in a pure state, see \eqref{solution}, the
Wigner ellipse has the form displayed in Fig.~3. In the limit of large
$r$, its major axis $\alpha$ is $\alpha\approx \E^r$, and its minor
axis is $\beta\approx\E^{-r}$. In the case of interaction with an
environment, the modes are in a mixed state given by
\eqref{dm-decohered}. For large $r$, the axes of the Wigner ellipse
are then given by
\bdm
\alpha\approx \E^r\ , \quad \beta\approx\sqrt{\frac{\xi}{k}} \gg \E^{-r}.
\edm
This corresponds to the situation displayed in Fig.~6. 

\begin{figure}[h]
  \begin{center}
    \includegraphics[width=0.6\textwidth]{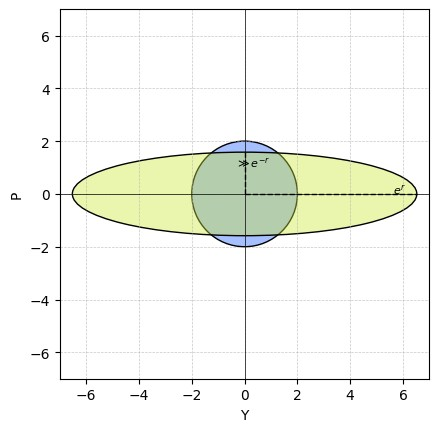}
  \caption[]{Due to decoherence, the ellipse (in green) becomes broader
    in $p$-direction, while the major axis remains approximately
    unchanged and alignes with the $y$-axis.}  
\end{center}
\end{figure}

The interaction of the primordial modes with other degrees of freedom
can, of course, not be so strong as to be in conflict with
observations. This means, in particular, that the correlation between
$y$ and $p$ cannot be completely washed out, because otherwise the
acoustic peaks in the CMB anisotropy spectrum observed by the \textsc{Planck}
satellite and other missions (Fig.~7) would not be present
\cite{KPS00}. In the formalism, this correlation condition is implemented by
\be
\lb{correlation}
\frac{\xi}{k}\ll\E^{2r}.
\ee

\begin{figure}[t]
  \begin{center}
    \includegraphics[width=1.0\textwidth]{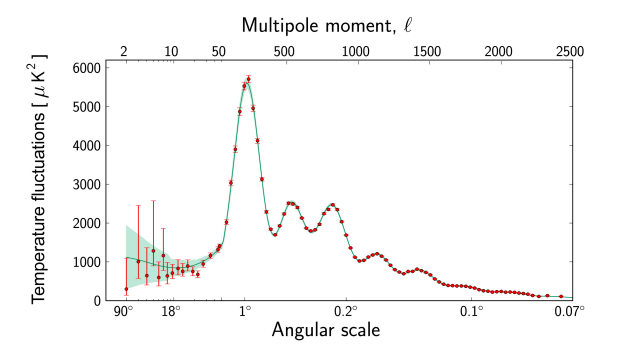}
  \caption[]{Power spectrum of temperature fluctuations in the Cosmic
    Microwave Background (CMB). \\ Copyright: \textsc{ESA} and the Planck
    Collaboration. See https://sci.esa.int/s/wRVmdjw.}  
\end{center}
\end{figure}

A related question is about the value of the entropy, $S$, for the
fluctuations. If correlations were fully absent (in contradiction to
observations), the entropy per mode would take for large $r$ the value
$S\approx 2r$. In our situation of pure entanglement,
Eq. \eqref{dm-decohered}, and remaining correlation
\eqref{correlation}, the entropy per mode turns out to be limited by
the value $S\approx r$, that is, only {\em half} of the maximum entropy $2r$
\cite{KPS00,KLPS07}. This corresponds to having the field-amplitude
basis as the pointer basis under decoherence (as opposed to e.g. the
particle-number basis).

The calculation of the entropy is based on the von Neumann formula
\bdm
S=-{\rm tr}(\rho_{\xi}\ln\rho_{\xi});
\edm
see \cite{KLPS07} for explicit calculations and expressions. Figure~8
shows the entropy in dependence of the dimensionless parameter
$\chi=:\xi/\Omega_{\rm R}$ controlling the strength of decoherence.
One recognizes that the asymptotic value for $\chi\gg1$ is readily
obtained; loss of one bit of information, which corresponds to
$S\approx\ln 2$, is obtained already for $\chi\approx 1.5$. 

\begin{figure}[t]
  \begin{center}
    \includegraphics[width=0.8\textwidth]{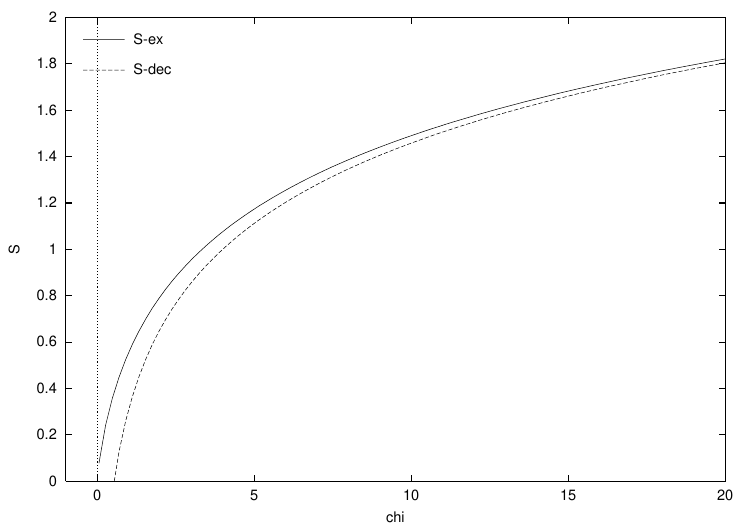}
  \caption[]{Entropy as a function of the decoherence parameter
    $\chi$. The solid curve is the exact expression, while the dotted
    curve gives the approximation for $\chi\gg 1$. Figure adapted from
    \cite{KLPS07}.}    
\end{center}
\end{figure}

The above considerations concern decoherence during inflation. Of
interest is also the question about the decohering
behaviour at later stages, in particular for the stage after the second
Hubble-crossing during the 
radiation- or matter-dominated phase, see Fig.~2. This was
investigated in our paper \cite{KLPS98}, where we could show that the
Wigner ellipse rotates slowly and that the correlation
between amplitude and momentum holds for a sufficiently long time. 

As a side remark, I want to mention that similar investigations can be
applied to the Hawking radiation for black holes \cite{CK01}. A
quantum field on the background of an object collapsing to a black
hole is also turned into a (two-mode) squeezed state. Interaction with
other degrees of freedom again leads to decoherence, but with a notable
difference to cosmology: here, the pointer basis is the particle-number
basis, leading to the maximal entropy associated with a thermal state --
the thermal state usually assumed for Hawking radiation. In view of
this, there is no reason to impose an information-loss ``paradox'', at
least not at the semiclassical level.\footnote{A scenario for the full
  quantum evolution of balck holes must, of course, come from a fundamental theory of
  quantum gravity.} 

The above considerations can be put on a solid formal basis by using
established methods from the quantum theory of open systems. Instead
of making a case-by case analysis of different interaction
Hamiltonians, it is more convenient to study master equations and
their solutions. In the simplest case of an ideal interaction similar to
\eqref{dm-decohered}, one has for a quantum-mechanical system (e.g. a
dust particle) in interaction with its environment (e.g. photons or
air molecules) the following master equation for the reduced density
matrix \cite{JZ85}:
\be
\lb{master}
\frac{\D\hat{\rho}}{\D t}=-\frac{\I}{2m\hbar}[\hat{p}^2,\hat{\rho}]
-\Lambda[\hat{x},[\hat{x},\hat{\rho}(t)]],
\ee
where $\Lambda$ encodes the details of the interaction (e.g. Thomson
cross section). Asymptotically, for large times, the density matrix
becomes diagonal in position space. An interstesting
property for the Wigner function was proven in \cite{DK02}: it
becomes strictly {\em positive}
after a certain finite time $t_{\rm d}=\sqrt{m/2\hbar\Lambda} $,
independent of the initial state. It is straightforward to add further
terms to \eqref{master} in order to include back action on the
system. In that case, the density matrix does not become exactly
diagonal, but a finite coherence length in position space remains
\cite{JZ85}. 

A general master equation encompassing many particular interactions is
the Lindblad equation for a Markovian (local in time) evolution,
\bdm
\hbar\frac{\partial\hat{\rho}}{\partial t}=-\I[\hat{H},\hat{\rho}]
+\hat{L}\hat{\rho}\hat{L}^{\dagger}-\frac12\hat{L}^{\dagger}\hat{L}\hat{\rho}
-\frac12\hat{\rho}\hat{L}^{\dagger}\hat{L}.
\edm
For the primordial fluctuations, this equation was discussed in detail
in our paper \cite{KLPS07}. In accordance with our presentation above,
we found that the pointer basis is given by the field-amplitude basis
and that the time scale for decoherence during inflation is given by
$H_{\rm I}^{-1}$. Results were also found for the decoherence time 
during radiation dominance outside the Hubble schale and for the
situation after the second crossing of the Hubble scale; 
see also \cite{KP09} for a review. 

The quantum-to-classical transition for primordial fluctuations (both
scalar and tensorial) continues to be of interest; see, for example,
\cite{BHKMV23} and the references therein. One might wonder,
for example, whether one could perform Bell-type ``experiments''
(observations) for the CMB in order to find possible violations of
Bell inequalities and thus prove directly the quantum origin of the
fluctuations. For modes being in a two-modes squeezed state, such a
violation indeed occurs, but because of decoherence it seems
impossible to observe such violations in the CMB \cite{MV17}. 

In Sec.~3 above, we have mentioned the feature of quantum Darwinism by
which information about the pointer states can be obtained from
observations of the environmental degrees of freedom \cite{WZ22}. It
would be of particular interest to study quantum Darwinism in
cosmology, for example through the observation of primordial
gravitational waves by the \textsc{LISA} mission. Whether this is
possible, only the future will show.

\section{Arrow of time}
\label{arrow}

From the fundamental point of view of a full quantum theory of gravity, all
degrees of freedom in Nature are described by quantum
theory. Classical behaviour is an emergent feature only and can be
understood by the process of decoherence. There is, in fact, a whole
hierarchy of classicality \cite{deco}. Starting from a full quantum
state of gravity and matter, one can understand how {\em first} the
classical appearance of spacetime emerges and {\em then}
the classical appearance of other `background
fields' such as the homogeneous part of the inflaton field
\cite{CK87}. Given that, the next level in the hierarchy of classicality 
comes from the decoherence of primordial
fluctuations discussed above. In this
sense, all the structure that we observe in our Universe (such as the
one seen in Fig.~1) can be understood as emerging from a quantum
origin. The picture is similar to the picture of a hypothetical inside view within
a nucleus as described in Sec.~3  -- but for the Universe, we can only
have an inside view. 

Decoherence is also important for an understanding of the arrow of
time, that is, for the origin of irreversibility in our
Universe. Alexei has for a long time entertained the idea
\cite{personal} that the
arrow of time should emerge from a breaking of the primordial symmetry
encoded in the hypothetical de~Sitter state assumed in his pioneering
paper \cite{AS79}. 
In \cite{Weyl}, I have made the proposal that the origin of
irreversibility can be traced back to a quantum version of Roger
Penrose's Weyl curvature hypothesis. In this, a crucial role is played
by assuming the initial state \eqref{initial} for the primordial
fluctuations, in particular for the tensor perturbations describing
primordial gravitational waves. With such an initial condition, an
initial unentangled state between degrees of freedom will evolve into
an entangled state, with an increase of entanglement entropy
\cite{KCT24}. This increasing entropy may eventually lead to the
Second Law and to the origin of probabilities according to the Born
rule \cite{CKT25}, although the final word on this can only be spoken
after a fundamental quantum theory of all interactions will be
established (if this will ever be the case \cite{CK24}). 

In connection with Alexei's suggestion of a symmetric initial state in
Sec.~2, I have mentioned the pre-Socratic philosophers Anaximander and
Anaxagoras. At least according to Aristotle in his {\em Eudemian
  Ethics}, Anaxagoras has answered to the question whether there is a
reason why someone should decide to be borne instead of remaining
unborne, that the reason is ``to observe the celestial edifice and the order in the
cosmos.'' Few people in our time have contributed so much to our
understanding of the cosmos than Alexei Starobinsky. We greatly miss him.

\begin{acknowledgement}
I am most grateful to the late Alexei Starobinsky for collaboration,
discussions, and many wonderful personal meetings that I will never
forget. I also thank Andrei Barvinsky, Alexander Kamenshchik, and
David Polarski for our enjoying collaboration on decoherence and other
fascinating topics.
\end{acknowledgement}
\ethics{Competing Interests}
{The author has no conflicts of interest to declare that are relevant
to the content of this chapter.}

\end{document}